\documentclass[12pt]{article}
\usepackage{times}

\usepackage{epsf}
\def\beq{\begin{equation}}
\def\eeq{\end{equation}}
\def\bea{\begin{eqnarray}}
\def\eea{\end{eqnarray}}

\def\vel{\left|}
\def\ver{\right|}
\def\nnb{\nonumber}

\def\rar{\rightarrow}
\def\nnb{\nonumber}

\def\ba{\begin{array}}
\def\ea{\end{array}}
\def\bea{\begin{eqnarray}}
\def\eea{\end{eqnarray}}

\def\Bell{$B_d\rightarrow \,\,\eta\,\, \ell^+ \ell^-$}
\def\Bepell{$B_{d}\rightarrow \,\,{\eta^{(\prime)}}\,\, \ell^+ \ell^-$}
\def\Bepett{$B_{d}\rightarrow \,\,{\eta}\,\, \tau^+ \tau^-$}
\def\Bepemm{$B_{d}\rightarrow \,\,{\eta}\,\, \mu^+ \mu^-$}

\def\tep{$b \rar d \ell^+ \ell^-$}

\title{The analysis of $B_{d}\rightarrow (\eta, \eta^{\prime})\ell^+\ell^-$ decays in the standard model}
\author{\vspace{1cm}\\
         {\bf G\"{u}ray Erkol}
         \thanks{E-mail: erkol@kvi.nl}\\\\ KVI, University of Groningen,\\ Zernikelaan 25, 9747 AA Groningen\\
		  The Netherlands \\ \\
        {\bf G\"{u}rsevil  Turan}
        \thanks{E-mail:
        gsevgur@metu.edu.tr}\\\\ Physics Department, Middle East Technical  University,\\
        Inonu Bulvari 06531, Ankara \\ Turkey \,\,}
\date{}
\begin{document}
\setlength{\baselineskip}{24pt} \maketitle
\setlength{\baselineskip}{4mm}

\abstract{We study the differential branching ratio, branching
ratio and the CP-violating asymmetry for the exclusive $B_{d}\rar (\eta, \eta^{\prime})\ell^+\ell^-$ 
decays in the standard model. We deduce the $B_{d}\rar (\eta, \eta^{\prime})$ form factors from the form 
factors of $B \rar \pi$ available in the literature, by using the $SU(3)_F$ symmetry. We observe that these 
decay modes, which are within the reach of forthcoming B-factories, are very promising to observe CP-violation.}
\thispagestyle{empty} \setcounter{page}{1}
\newpage
\section{Introduction \label{s1}}
The decays of B-meson are very promising for investigating the Standard Model (SM) and searching for the new 
physics beyond it. Among these B-decays, the rare semileptonic ones have attracted much attention for a long
time, since  they offer the most direct methods to determine the weak mixing angles 
and Cabibbo-Kobayashi-Maskawa (CKM) matrix elements. These decays can also be very useful to test the various
new physics scenarios like the two Higgs doublet models (2HDM), 
minimal supersymmetric standard model (MSSM)\cite{MSSM}. etc.

From the experimental side, there is an impressive effort to search for B-decays, in B-factories such 
as Belle, BaBar, LHC-B. For example, CLEO Collaboration reports for the branching ratios $(BRs)$ of the
$B^0\rar\pi^-\ell^+\nu$ and $B^0\rar\rho^-\ell^+\nu$ decays \cite{Cleo1} as
\bea
BR(B^0\rar\pi^-\ell^+\nu)&=&(1.8\pm 0.4\pm 0.3 \pm 0.2)\times 10^{-4}\nnb\\
BR(B^0\rar\rho^-\ell^+\nu)&=&(2.57\pm 0.29_{-0.46}^{+0.33}\pm 0.41) \times 10^{-4}.
\eea
From these results, the value of the CKM matrix element $|V_{ub}|=3.25 \pm 0.14^{+0.21}_{-0.29} \pm 0.55$ has 
been determined \cite{Cleo1}. Recently, the BR of the inclusive $B\rar X_s \ell^+\ell^-$ decay has 
been also reported by Belle Collaboration \cite{Belle2};
\bea
BR(B\rar X_s \ell^+\ell^-)=(6.1\pm 1.4^{+1.4}_{-1.1}) \times 10^{-6}\, ,
\eea
which is very close to the value predicted by the SM \cite{Ali0}.

In this paper, we investigate the \Bepell decay modes within the SM. 
It is well known that the inclusive rare decays are more difficult to measure, although 
they are theoretically cleaner than the exclusive ones. This motivates the study of exclusive decays, but 
their theoretical investigation requires the additional knowledge of decay form factors, i.e. the matrix 
elements of the effective Hamiltonian between initial B and the final meson states. The nonperturbative 
sector of QCD is used in order to determine these form factors. Two of the form factors, $f_+$ and $f_-$,
necessary for \Bell decay have been calculated very recently, in the framework of light-cone QCD sum 
rules  \cite{Aliev5}. However, we do not have a precise calculation on the remaining form factor, 
$f_T$ for \Bepell decay yet. Therefore, in this work, we choose to deduce the form factors of 
$B_d\rightarrow\eta^{(\prime)}$ transition  from the form factors of $B\rar \pi$ using 
 the $SU(3)_F$ symmetry. The form factors of $B\rar\pi$ have been 
 calculated in  the light-cone constituent quark model (LCQM) \cite{Melikhov1,Melikhov2} and also in the 
 QCD sum rules method  (QCDSR) \cite{Ball}; and in this paper, we will give our numerical results using  
 both of these approaches. Let us mension that the $B\rar K$ hadronic matrix elements computed in LCQM and QCDSR have been used to evaluate the semileptonic rate of the $B\rar K \ell^+ \ell^-$ decay mode \cite{BtoK1, BtoK2}. Compared to the recently measured value of $BR(B\rar K \ell^+ \ell^-)=(0.75^{+0.25}_{-0.21}\pm 0.09)\times 10^{-6}$ by Belle Collaboration \cite{KEK1} and also BaBar Collaboration \cite{BABAR1}, we see that QCDSR predicts a better result.

In this work, we also calculate the CP asymmetry in the \Bepell decay, which is induced by the \tep transition 
 at the quark level. For $b\rightarrow s\ell^+\ell^-$ 
transition, the matrix element contains the terms that receive
contributions from $t\bar{t}$, $c\bar{c}$ and $u\bar{u}$ loops,
which are proportional to the combination of
$\xi_t=V_{tb}V^*_{ts}$, $\xi_c=V_{cb}V^*_{cs}$ and
$\xi_u=V_{ub}V^*_{us}$, respectively. Smallness of $\xi_u$
in comparison with $\xi_c$ and $\xi_t$, together with the
unitarity of the CKM matrix elements, bring about the consequence
that matrix element for the $b \rightarrow s\ell^+\ell^-$ decay
involves only one independent CKM factor $\xi_t$, so that the
CP violation in this channel is suppressed  in the SM \cite{Aliev00,Du}.
However, for $b \rightarrow d \ell^+\ell^-$ decay, all the CKM
factors $\eta_t=V_{tb}V^*_{td}$, $\eta_c=V_{cb}V^*_{cd}$ and
$\eta_u=V_{ub}V^*_{ud}$ are at the same order in the SM so that
they can induce a CP violating asymmetry between the decay rates of
the reactions $b \rightarrow d \ell^+\ell^-$  and $\bar{b}\rightarrow \bar{d}\ell^+\ell^-$ 
\cite{Kruger2}. So, $b \rightarrow d \ell^+\ell^-$ decay seems to be suitable 
for establishing CP violation in B mesons. On the other hand, it should be 
noted that the detection of the $b \rightarrow d \ell^+\ell^-$ decay
will probably be more difficult in the presence of a much stronger decay 
$b \rightarrow s \ell^+\ell^-$ and this would make the corresponding exclusive
decay channels more preferable in search of CP violation. In this context, the exclusive
$B_{d} \rar (\pi,\rho) \, \ell^+ \ell^-$, and $B_{d} \rar \gamma \, \ell^+ \ell^-$ decays 
have been extensively  studied in the SM \cite{Kruger1, Erkol} and beyond \cite{Aliev4}-\cite{Choud2}.

The paper is organized as follows: In section \ref{sec1}, first the effective 
Hamiltonian is presented and the form factors are defined. Then, the basic formulas
of the differential  branching ratio dBR/ds, branching ratio $BR$ and the CP violating asymmetry
 $A_{CP}$ for \Bepell decays are introduced. Section \ref{sec2} is devoted to the numerical 
 analysis and discussion.

\section{Effective Hamiltonian and Form Factors}\label{sec1}
The leading order QCD corrected effective
Hamiltonian, which is induced by the corresponding quark level
process $b \rar d \,  \ell^+ \ell^-$, is given by
\cite{Buchalla}-\cite{Misiak}:
\begin{eqnarray}\label{Hamiltonian} {\cal H}_{eff}  =  \frac{4
G_F}{\sqrt{2}} \, V_{tb} V^*_{td}\Bigg\{ \sum_{i=1}^{10}& \, \,
C_i (\mu ) \, O_i(\mu)-\lambda_u
\{C_1(\mu)[O_1^u(\mu)-O_1(\mu)]\nnb\\&+C_2(\mu)[O_2^u(\mu)-O_2(\mu)]\}\Bigg\}
\end{eqnarray}
where 
\bea\label{CKM}
\lambda_u=\frac{V_{ub}V_{ud}^\ast}{V_{tb}V_{td}^\ast}, 
\eea 
using the unitarity of the CKM matrix i.e.
$V_{tb}V_{td}^\ast+V_{ub}V_{ud}^\ast=-V_{cb}V_{cd}^\ast$. The
explicit forms of the operators $O_i$ can be found in
refs. \cite{Buchalla,Wise}. In Eq.(\ref{Hamiltonian}),
$C_i(\mu)$ are the Wilson coefficients calculated at a
renormalization point $\mu$ and their evolution from the higher scale $\mu=m_W$
down to the low-energy scale $\mu=m_b$ is described by the renormalization group
equation. For $C^{eff}_7(\mu)$ this calculation is performed in refs.\cite{Borzumati,Ciu}
upto next to leading order. The value of $C_{10}(m_b)$ to the leading logarithmic approximation 
can be found e.g. in
\cite{Buchalla,Misiak}. The terms that are the source of the CP violation are given
by the following, which have a perturbative
part and a part coming from long distance (LD) effects due to conversion of the
real $\bar{c}c$ into lepton pair $\ell^+ \ell^-$:
\begin{eqnarray}
C_9^{eff}(\mu)=C_9^{pert}(\mu)+ Y_{reson}(s)\,\, ,
\label{C9efftot}
\end{eqnarray}
where
\begin{eqnarray}\label{Cpert}
C_9^{pert}(\mu)&=& C_{9}+h(u,s) [ 3 C_1(\mu) + C_2(\mu) + 3
C_3(\mu) + C_4(\mu) + 3 C_5(\mu) + C_6(\mu) \nonumber
\\&+&\lambda_u(3C_1 + C_2) ] -  \frac{1}{2} h(1, s) \left( 4
C_3(\mu) + 4 C_4(\mu)
+ 3 C_5(\mu) + C_6(\mu) \right)\nnb \\
&- &  \frac{1}{2} h(0,  s) \left[ C_3(\mu) + 3 C_4(\mu) +\lambda_u
(6 C_1(\mu)+2C_2(\mu)) \right] \\&+& \frac{2}{9} \left( 3 C_3(\mu)
+ C_4(\mu) + 3 C_5(\mu) + C_6(\mu) \right) \nonumber \,\, ,
\end{eqnarray}
and
\begin{eqnarray}
Y_{reson}(s)&=&-\frac{3}{\alpha^2_{em}}\kappa \sum_{V_i=\psi_i}
\frac{\pi \Gamma(V_i\rightarrow \ell^+
\ell^-)m_{V_i}}{m_B^2 s-m_{V_i}+i m_{V_i}
\Gamma_{V_i}} \nonumber \\
&\times & [ (3 C_1(\mu) + C_2(\mu) + 3 C_3(\mu) + C_4(\mu) + 3
C_5(\mu) + C_6(\mu))\nnb\\ &+&\lambda_u(3C_1(\mu)+C_2(\mu))]\, .
 \label{Yresx}
\end{eqnarray}
In Eq.(\ref{Cpert}), $s=q^2/m_B^2$ where $q$ is the momentum transfer, $u=\frac{m_c}{m_b}$
 and the functions $h(u, s)$ arise from one loop 
contributions of the four-quark operators $O_1-O_6$ and are given by
\begin{eqnarray}
h(u, s) &=& -\frac{8}{9}\ln\frac{m_b}{\mu} - \frac{8}{9}\ln u +
\frac{8}{27} + \frac{4}{9} y \\
& & - \frac{2}{9} (2+y) |1-y|^{1/2} \left\{\begin{array}{ll}
\left( \ln\left| \frac{\sqrt{1-y} + 1}{\sqrt{1-y} - 1}\right| -
i\pi \right), &\mbox{for } y \equiv \frac{4u^2}{ s} < 1 \nonumber \\
2 \arctan \frac{1}{\sqrt{y-1}}, & \mbox{for } y \equiv \frac
{4u^2}{ s} > 1,
\end{array}
\right. \\
h(0,s) &=& \frac{8}{27} -\frac{8}{9} \ln\frac{m_b}{\mu} -
\frac{4}{9} \ln s + \frac{4}{9} i\pi \,\, . \label{hfunc}
\end{eqnarray}
The phenomenological parameter $\kappa$
in Eq. (\ref{Yresx}) is taken as $2.3$ (see e.g., \cite{Kruger2}).  

Neglecting the mass of the $d$ quark, the effective short distance Hamiltonian
for the $b \rightarrow d \ell^+ \ell^-$ decay in Eq.(\ref{Hamiltonian})  leads to the QCD
corrected matrix element:
\begin{eqnarray}\label{genmatrix}
{\cal M} &=&\frac{G_{F}\alpha}{2\sqrt{2}\pi }V_{tb}V_{td}^{\ast }%
\Bigg\{C_{9}^{eff}(m_{b})~\bar{d}\gamma _{\mu }(1-\gamma _{5})b\,\bar{\ell}%
\gamma ^{\mu }\ell +C_{10}(m_{b})~\bar{d}\gamma _{\mu }(1-\gamma _{5})b\,\bar{%
\ell}\gamma ^{\mu }\gamma _{5}\ell  \nonumber \\
&-&2C_{7}^{eff}(m_{b})~\frac{m_{b}}{q^{2}}\bar{d}i\sigma _{\mu \nu
}q^{\nu }(1+\gamma _{5})b\,\bar{\ell}\gamma ^{\mu }\ell
\Bigg\}.\nonumber\\
\end{eqnarray}

Next we proceed to calculate the  $BR$s of the \Bepell
decays. The necessary matrix elements to do this are
$<{{\eta^{(\prime)}}}(p_{{\eta^{(\prime)}}})|\bar{d}\gamma_\mu(1-\gamma_5)b|B(p_B)>$,
$<{{\eta^{(\prime)}}}(p_{{\eta^{(\prime)}}})|\bar{d}i\sigma_{\mu\nu}q_\nu(1+\gamma_5)b|B(p_B)>$ and
$<{{\eta^{(\prime)}}}(p_{{\eta^{(\prime)}}})|\bar{d}(1+\gamma_5)b|B(p_B)>$.
The first two of these matrix elements can be written in terms of the form factors
in the following way

\bea\label{matrixp1}
<{{\eta^{(\prime)}}}(p_{{\eta^{(\prime)}}})|\bar{d}\gamma_\mu(1-\gamma_5)b|B(p_B)>&=&f^+(q^2)(p_B+p_{{\eta^{(\prime)}}})_\mu
+f^-(q^2)q_\mu\, ,\eea \bea\label{matrixp2}
<{{\eta^{(\prime)}}}(p_{{\eta^{(\prime)}}})|\bar{d}i\sigma_{\mu\nu}q^\nu(1+\gamma_5)b|B(p_B)>& \!\!\!=\!\!\!
&[(p_B+p_{\eta^{(\prime
)}})_\mu
q^2-q_\mu(m_B^2-{m_{\eta^{(\prime)}}}^2)]f_v(q^2)\, ,
\eea
where $p_B$ and $p_{{\eta^{(\prime)}}}$ denote the four momentum vectors of $B$
and ${\eta^{(\prime)}}$-mesons, respectively. $f_v(q^2)$ is sometimes written as $f_v(q^2)=f_T/(m_B+m_{\eta^{(\prime)}}^2)$.

To find $<{{\eta^{(\prime)}}}(p_{{\eta^{(\prime)}}})|\bar{d}(1+\gamma_5)b|B(p_B)>$, we multiply both sides
of  Eq. (\ref{matrixp1}) with  $q_{\mu}$ and then use the equation
of motion. Neglecting the mass of the $d$-quark, we get
\bea\label{matrixp3}
<{{\eta^{(\prime)}}}(p_{\eta^{(\prime)}})|\bar{d}(1+\gamma_5)b|B(p_B)>=\frac{1}{m_b}[f^+(q^2)(m_B^2-{m_{{\eta^{(\prime
)}}}}^2)+f^-(q^2)
q^2 ].
\eea

As pointed out in sec.1, although the form factors $f_+$ and $f_-$ for $B\rightarrow \eta$ decay
have  been calculated in the framework of the
light-cone QCD sum rules in  \cite{Aliev5}, we do not have a precise calculation of the other form 
factor $f_v$ in the literature yet. However, the form factors of $B_d\rightarrow\eta^{(\prime)}$ transition
can be related  to those of $B\rar\pi$   through the  $SU(3)_F$ symmetry \cite{Kim, Skands}. In addition, 
the authors of \cite{Aliev5} emphasize that, their results coincide with the ones that are calculated 
using the $SU(3)_F$ symmetry. Therefore,  we choose to deduce the form factors necessary in this work
from the  $B\rar \pi$ transition using the $SU(3)_F$ symmetry. For $\eta-\eta^\prime$ mixing, 
we adopt the following scheme \cite{eta, eta2},

\bea\label{mixing}
|\eta>&=&\cos \,\phi |\eta_q>\,-\,\sin\,\phi |\eta_s>, \nnb\\
|\eta^\prime>&=&\sin \,\phi |\eta_q>\,+\, \cos\,\phi |\eta_s>,
\eea
where $|\eta_q>=(u\bar{u}+d\bar{d})/\sqrt{2}$, $|\eta_s>=s\bar{s}$, and $\phi=39.3$ is the fitted mixing angle \cite{eta}. Hence, the relation between the form factors are written as follows:

\bea
F^{B_d\rar\eta}(q^2)&=&\cos\,\phi\,F^{B\rar\pi}(q^2),\nnb\\
F^{B_d\rar\eta^\prime}(q^2)&=&\sin\,\phi\,F^{B\rar\pi}(q^2).
\eea
For $B\rightarrow \pi$, we use the results calculated in two different frameworks: In the LCQM, the form
factors are parametrized in the following pole forms \cite{Melikhov1,Melikhov2} 
\bea
f^+ (q^2) &=& \frac{0.29}{\left(1 - \frac{q^2}{6.71^2}\right)^{2.35}}~, 
f^-(q^2) = - \frac{0.26}{\left(1 - \frac{q^2}{6.553^2}\right)^{2.30}}~, \nnb \\
f_v(q^2) &=& - \frac{0.05}{\left(1-\frac{q^2}{6.68}\right)^{2.31}}~.\label{pfm}
\eea
However in the QCDSR approach, they are given by \cite{Ball}
\bea
f^+ (q^2) &=& \frac{0.305}{\left(1 - 1.29 \frac{q^2}{m_B^2}+0.206(\frac{q^2}{m_B^2})^2\right)}~, 
f_0(q^2) =  \frac{0.305}{\left(1-0.266 \frac{q^2}{m_B^2}-0.752(\frac{q^2}{m_B^2})^2\right)}~, \nnb \\
f_T(q^2) &=&  \frac{0.296}{\left(1-1.28\frac{q^2}{m_B^2}+0.193(\frac{q^2}{m_B^2})^2\right)}~,\label{pfm2}
\eea
from which $f^-$ can be calculated through the relation:
\bea 
f^{-}=(m_B^2-m_{\eta^{(\prime)}}^2)(f_0-f^+ )/q^2.
\eea
Using the above matrix elements,
we find the amplitudes governing the \Bepell   decays as follows:

\bea\label{matrixp} \cal{M}^{B\rightarrow \eta^{(\prime)}
}&=&\frac{G_F\alpha}{2\sqrt{2}\pi}V_{tb}V_{td}^\ast\Bigg \{[2A
p_{{{\eta^{(\prime)}}}}^\mu+B q^\mu]\bar{\ell}\gamma_\mu \ell+[2G p_{{{\eta^{(\prime)}}}}^\mu +D
q^\mu]\bar{\ell}\gamma_\mu\gamma_5\ell \Bigg\}~,\nnb\\
\eea
where
\bea
A&=&C^{eff}_9 f^+ -2m_B C^{eff}_7 f_v,\nnb\\
B&=&C^{eff}_9(f^+ +f^-)+2 C^{eff}_7
\frac{m_B}{q^2}f_v(m_B^2-m_{{{\eta^{(\prime)}}}}^2-q^2),\nnb\\
G&=&C_{10}f^+,\nnb\\
D&=&C_{10}(f^+ +f^-).
\eea

Using Eq.(\ref{matrixp}) and performing summation over final  lepton polarization, we get for the double
differential decay rates:

\begin{eqnarray}\label{ddrp}
\frac{d^2\Gamma^{B\rightarrow {{\eta^{(\prime)}}}}}{ds ~dz}&=&\frac{G_F^2
\alpha^2}{2^{11}\pi^5}|V_{tb}V_{td}^\ast|^2 m_B^3
\sqrt{\lambda}~v~ \Bigg\{m_B^2 \lambda(1-z^2 v^2)|A|^2 \nonumber
\\&+& (m_B^2 \lambda(1-z^2 v^2) +16~r~m^2_{\ell})~|G|^2
+4~s~m_{\ell}^2~|D|^2\nonumber\\
&+&
4~m_{\ell}^2~(1-r-s)~Re[G\,D^\ast]\Bigg\}
\, ,\end{eqnarray}

Here $s=q^2/m_B^2$, $r=m_{{{\eta^{(\prime)}}}}^2/m_B^2$,
$v=\sqrt{1-\frac{4t^2}{s}}$, $t=m_\ell^2/m_B^2$, 
$\lambda=r^2+(s-1)^2-2r(s+1)$, and
$z=\cos\theta$, where $\theta$ is the angle between the
three-momentum of the $\ell^-$ lepton and that of the B-meson in
the center of mass frame of the dileptons $\ell^+\ell^-$. After integrating over the angle variable we find
\bea
\frac{d\Gamma^{B\rightarrow {{\eta^{(\prime)}}}}}{ds}&=&\frac{G_F^2
\alpha^2}{2^{10}\pi^5}|V_{tb}V_{td}^\ast|^2 m_B^3 \sqrt{\lambda}
\, v \, \Delta \, , \eea where \bea \Delta & = &
\frac{1}{3}~m_B^2~\lambda(3-v^2)(|A|^2+|G|^2)
+\frac{4m_{\ell}^2}{3s}(12\,r~s+\lambda)|G|^2\nnb\\&+&4~m_{\ell}^2~s~|D|^2+4~m_{\ell}^2 (1-r-s) Re[G~D^\ast]\,.
\label{deltapi} \eea

We now consider the CP violating asymmetry, $A_{CP}$, between the \Bepell and
$\overline{B}_{d}\rightarrow \,\,\overline{\eta}^{(\prime)}\,\, \ell^+ \ell^-$ decays, which  is defined 
as follows:
\bea A_{CP}(x) =  \frac{\Gamma (B_{d}\rightarrow \,\,\eta^{(\prime)}\,\, \ell^+ \ell^-)
-\Gamma (\overline{B}_{d}\rightarrow \,\,\overline{\eta}^{(\prime)}\,\, \ell^+ \ell^-) }
{\Gamma (B_{d}\rightarrow \,\,{{\eta^{(\prime)}}}\,\, \ell^+ \ell^-) + 
\Gamma (\overline{B}_{d}\rightarrow \,\,\overline{\eta}^{(\prime)}\,\, \ell^+ \ell^-)} ~~. \label{ACP1}
\end{eqnarray}
Using this definition we calculate the $A_{CP}$ as: \bea
A_{CP}=\frac{\int H(s)~ds}{\int (\Delta-H(s))~ds},\eea where
\bea
H(s)&=&\frac{2}{3}\,f_+ m_B^2(3-\upsilon^2)\lambda~Im~\lambda_u~\Bigg( Im~\xi_2~ C_7^{eff}~f_T\frac{2m_b}{m_B+m_{{{\eta^{(\prime)}}}}}-f_+~(Im~\xi_1^\ast~\xi_2)\Bigg).
\eea In calculating
this expression, we use the following parametrizations: \bea
C_9^{eff}\equiv\xi_1+\lambda_u~\xi_2\, , \eea
\bea \label{lamu}\lambda_u=\frac{\rho(1-\rho)-\eta^2-i\eta}{(1-\rho)^2+\eta^2}+O(\lambda^2).\eea  

\section{Numerical Results and Discussion}\label{sec2}
 In this section we present the numerical results of our calculations related to 
 \Bepell $(\ell=e,\,\mu,\,\tau)$ 
decays,  for four different sets of parameter choice of the form factors and the updated fits of the Wolfenstein parameters \cite{Ali},
which are summarized in Table \ref{tab1}.
The total $BR$s are collected in Table \ref{tab2}.  
We have also evaluated the average values of CP asymmetry $<A_{CP}>$ 
in \Bepell decays for the above sets of parameters, and our results are
displayed in Table \ref{tab3}. In both tables, the values in the paranthesis are the corresponding quantities
calculated without including  the long distance effects. We observe that the results of $<A_{CP}>$
is very sensitive to the choice of four different sets of parameters for $\tau$ channel, while  they are 
very close to each other for $\mu$ channel.   
\begin{table}
\begin{center}
\begin{tabular}{|c|c|c|}
 \hline
                       & $(\rho ; \eta)$ & Form ~~factors           \\ \hline \hline
       set-1           & $(0.3;0.34)$    & {\small LCQM}                      \\ \hline
       set-2	       & $(0.15;0.34)$  & {\small LCQM}                      \\ \hline
       set-3           & $(0.3;0.34)$    & {\small QCDSR}                   \\ \hline
       set-4           & $(0.15;0.34)$  & {\small QCDSR}                   \\ \hline
  \end{tabular}
  \end{center}
  \caption{List of the values for the Wolfenstein parameters and the form factors of the 
  transition $B\rightarrow \pi $ calculated in the light-cone constituent quark model (LCQM) 
  \cite{Melikhov1,Melikhov2} and light-cone QCD sum rule approach (QCDSR) \cite{Ball}.}\label{tab1}
  \end{table}	

The input parameters and the initial values of the Wilson coefficients we used in our numerical
analysis are as follows:
\begin{eqnarray}
& & m_B =5.28 \, GeV \, , \, m_b =4.8 \, GeV \, , \,m_c =1.4 \,
GeV \, , m_{\tau} =1.78 \, GeV,\, \nnb \\
& &  m_{\mu}=0.105\,GeV,\, \vel V_{tb}V_{td}^\ast\ver=0.01,\,m_{\eta}=0.547\, GeV,\, m_{\eta^\prime}=0.958\, GeV, \,\nnb\\ 
& &C_1=-0.245,\, C_2=1.107,\, C_3=0.011,\, C_4=-0.026,\,C_5=0.007,\,\nnb\\
& & C_{6}=-0.0314,\,C^{eff}_{7}=-0.315,\, C_{9}=4.220, \, C_{10}=-4.619.
\end{eqnarray}

There are five possible resonances in the $c\bar{c}$ system that
can contribute to the decay under consideration and to calculate
their contributions, we need to divide the integration region for
$s$ into three parts for $\ell=e,\mu $ so that we have $4
m^2_{\ell}/m_B^2 \leq s \leq (m_{\psi_1}-0.02)^2/m_B^2$ and
$(m_{\psi_1}+0.02)^2/m_B^2 \leq s \leq (m_{\psi_2}-0.02)^2/m_B^2$ and
$(m_{\psi_2}+0.02)^2/m_B^2 \leq s \leq (m_B-m_{\eta^{(\prime)}})^2/m_B^2$, while for
$\ell=\tau$ it takes the form given by $4 m^2_{\tau}/m_B^2 \leq s \leq
(m_{\psi_2}-0.02)^2/m_B^2$ and $(m_{\psi_2}+0.02)^2/m_B^2 \leq s \leq
(m_B-m_{\eta^{(\prime)}})^2/m_B^2$ . Here, $m_{\psi_1}$ and $m_{\psi_2}$ are the
masses of the first and the second resonances, respectively. 
\begin{table}
\begin{center}
\begin{tabular}{|c|c|c|c|c|c|}
 \hline
  $10^{8} \cdot BR$    & {\scriptsize$\ell$} & set1          &  set2    & set3         &  set4 \\ \hline \hline
                       & $\tau$              & $0.331$       & $0.313$  & $0.687$      &  $0.659$    \\
		               &                     & $(0.324)$     & $(0.314)$& $(0.695)$    &  $(0.677)$ \\ \cline{2-6}
  $\eta$               & $\mu$               & $2.704$       & $2.511$  & $3.704$      &  $3.468$     \\
                       &                     & $(2.119)$     & $(2.063)$& $(3.049)$    &  $(2.966)$ \\ \cline{2-6}
                       & $e$                 & $2.713$       & $2.520$  & $3.716$      &  $3.479$    \\
					   &                     & $(2.127)$     & $(2.371)$& $(3.059)$    &  $(2.976)$  \\ \hline \hline
		               & $\tau$              & $0.092$       & $0.087$  & $0.153$      &  $0.146$    \\
		               &                     & $(0.086)$     & $(0.083)$& $(0.147)$    &  $(0.144)$ \\ \cline{2-6}
  $\eta'$              & $\mu$               & $1.363$       & $1.268$  & $1.779$      &  $1.666$     \\
                       &                     & $(1.033)$     & $(1.010)$& $(1.395)$    &  $(1.365)$ \\ \cline{2-6}
                       & $e$                 & $1.369$       & $1.273$  & $1.786$      &  $1.674$    \\
					   &                     & $(1.038)$     & $(1.015)$& $(1.402)$    &  $(1.372)$ \\ \hline
	\end{tabular}
  \end{center}
  \caption{The SM predictions for the integrated branching ratios for $\ell =\tau , \mu , e$ of the
  $B_d \rightarrow \eta^{(\prime)} \ell \ell$ decay with (without)  the long-distance effects.}\label{tab2}
  \end{table}	  

\begin{figure}[htb]
\vskip 0truein \centering \epsfxsize=3.8in
\leavevmode\epsffile{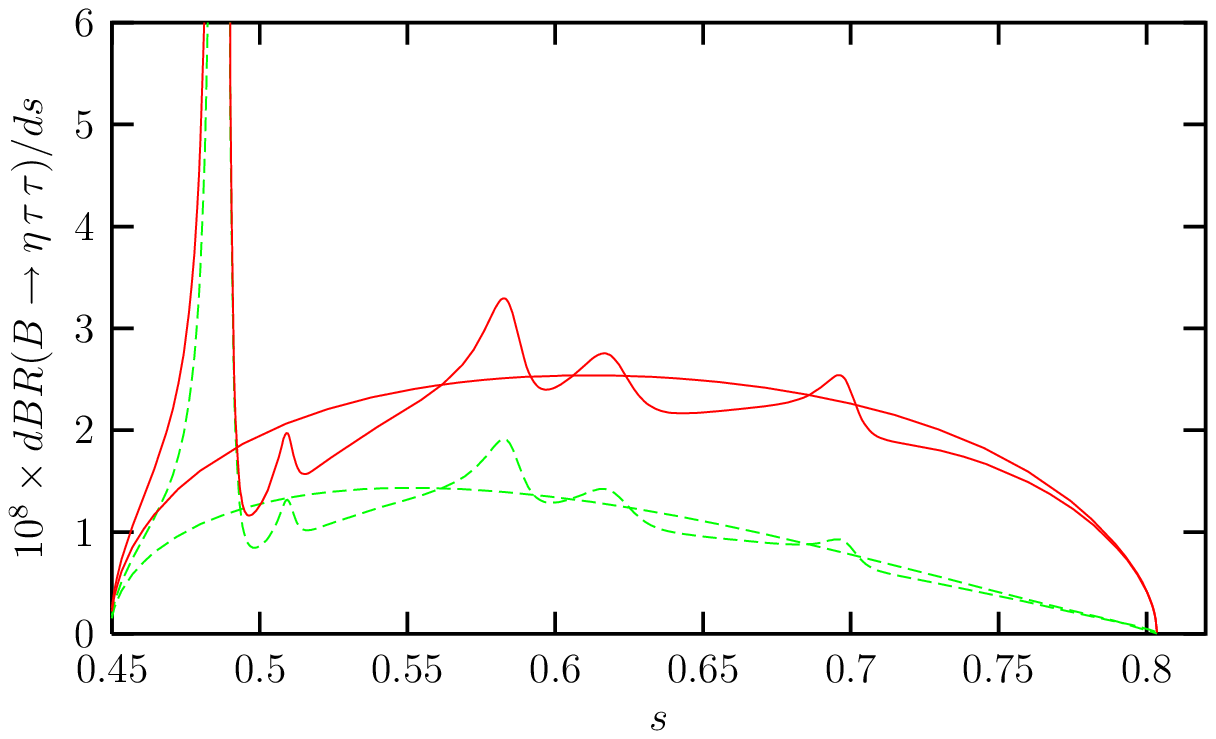} \vskip 0truein
\caption[]{Differential branching ratio for $B\rar \eta \, \tau^{+} \tau^{-}$ decay as a function of $s$
for the parameter set-1 and set-3,  represented by the dashed and the solid curves, respectively.
The sharp peaks in the figures are due to the long distance contributions.} \label{dBRetatau}
\end{figure}

\begin{figure}[htb]
\vskip 0truein \centering \epsfxsize=3.8in
\leavevmode\epsffile{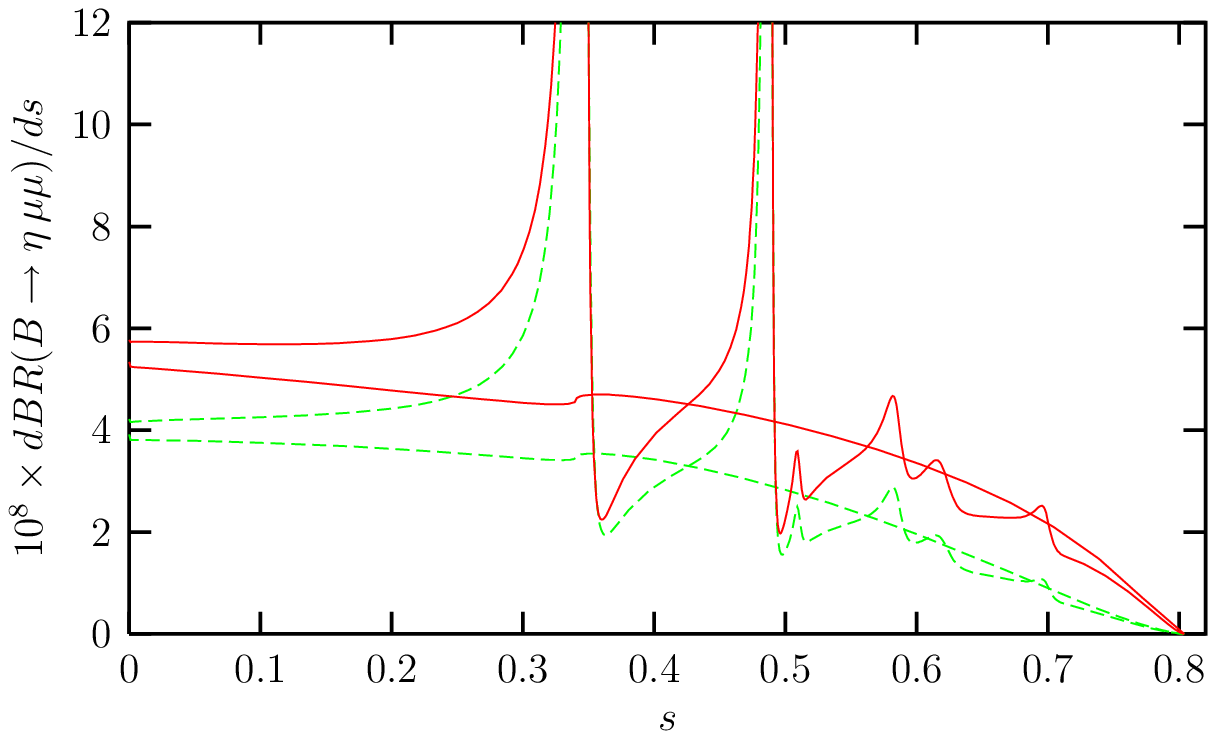} \vskip 0truein \caption[]{The same
as Fig.(\ref{dBRetatau}) but for the $B\rar \eta \, \mu^{+} \mu^{-}$ decay} \label{dBRetamu}
\end{figure}

\begin{figure}[htb]
\vskip 0truein \centering \epsfxsize=3.8in
\leavevmode\epsffile{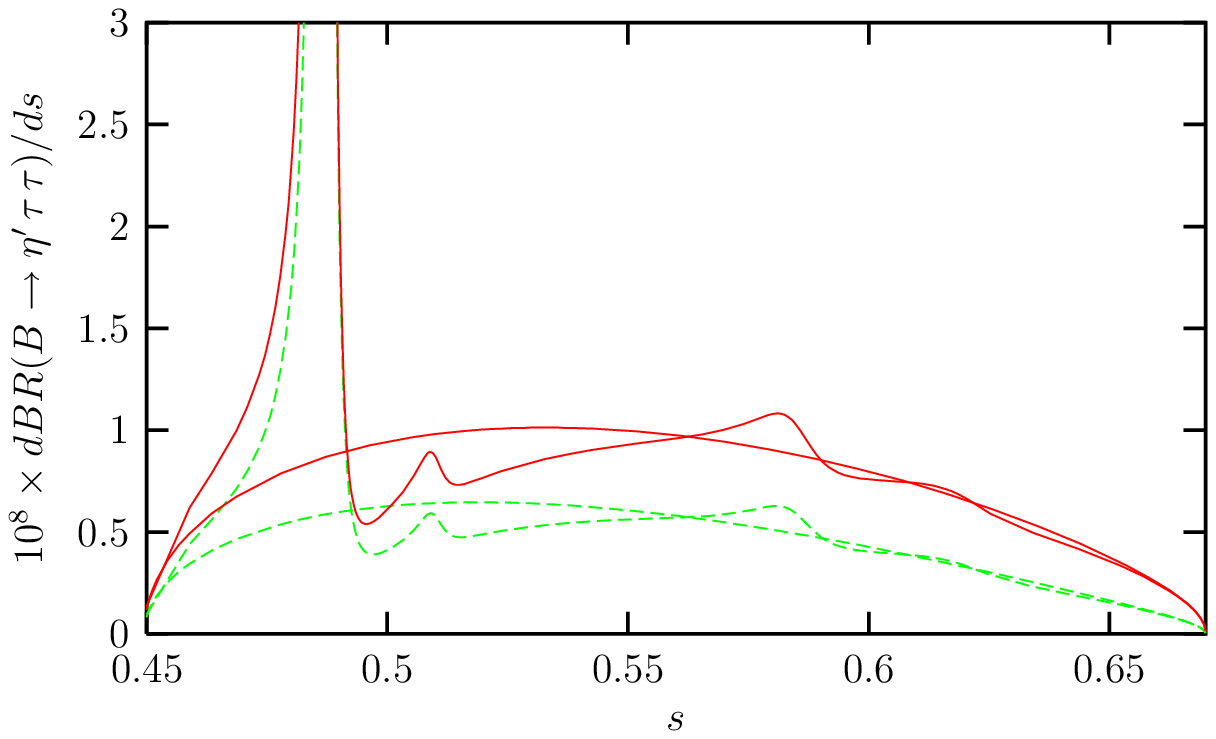} \vskip 0truein \caption[]{The same
as Fig.(\ref{dBRetatau}) but for the $B\rar \eta^{\prime} \, \tau^{+} \tau^{-}$ decay} \label{dBRetaptau}
\end{figure}

\begin{figure}[htb]
\vskip 0truein \centering \epsfxsize=3.8in
\leavevmode\epsffile{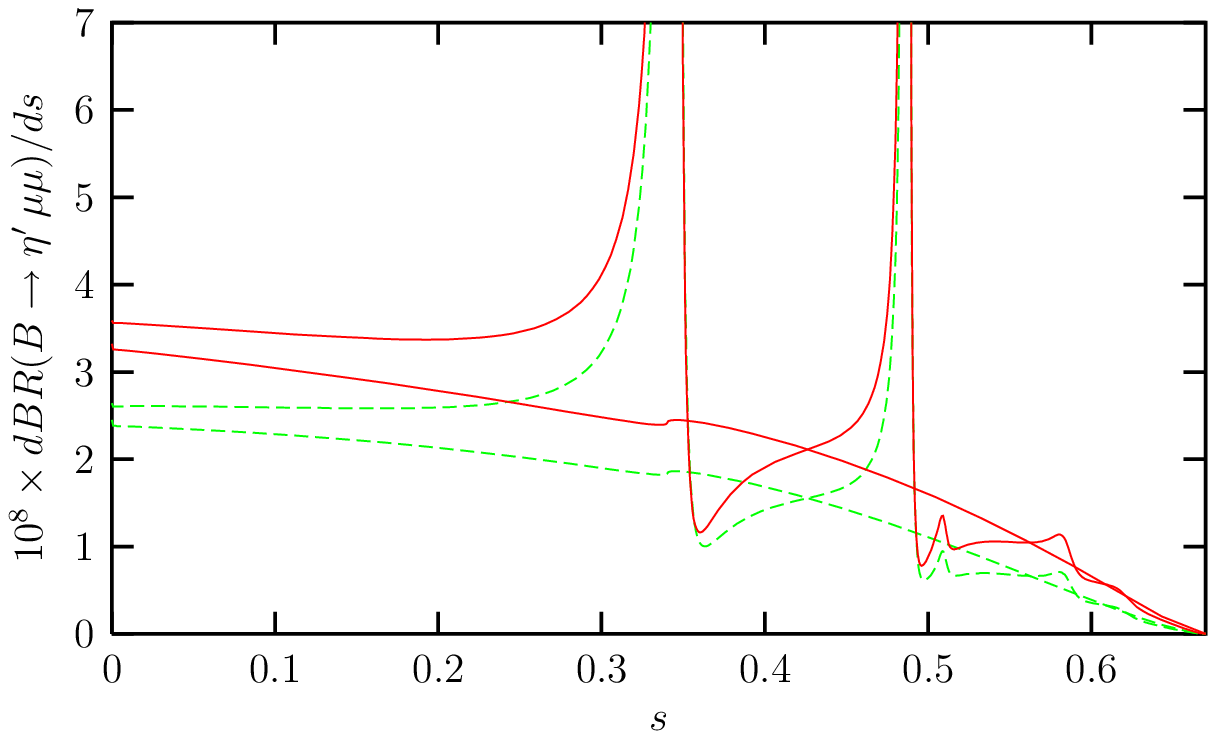} \vskip 0truein
\caption[]{The same
as Fig.(\ref{dBRetatau}) but for the $B\rar \eta^{\prime} \, \mu^{+} \mu^{-}$ decay} \label{dBRetapmu}
\end{figure}

In Fig. (\ref{dBRetatau}) and Fig. (\ref{dBRetamu}), we present the dependence of the $BR$ on 
the invariant mass of dileptons, $s$, for the \Bepett and  \Bepemm decays, respectively.
 We plot these graphs for the parameter set-1 and set-3
in Table \ref{tab1},  represented by the dashed and the solid curves, respectively.
The sharp peaks in the figures are due to the long distance contributions. As can be seen from 
these graphs, $BR$ stands more for the parameter set-3.
The same analysis above is made for $B_{d}\rightarrow \,\,{\eta^{\prime}}\,\, \tau^+ \tau^-$ and $B_{d}\rightarrow \,\,{\eta^{\prime}}\,\, \mu^+ \mu^-$ 
decays in Fig. (\ref{dBRetaptau}) and Fig. (\ref{dBRetapmu}), respectively.

\begin{figure}[htb]
\vskip 0truein \centering \epsfxsize=3.8in
\leavevmode\epsffile{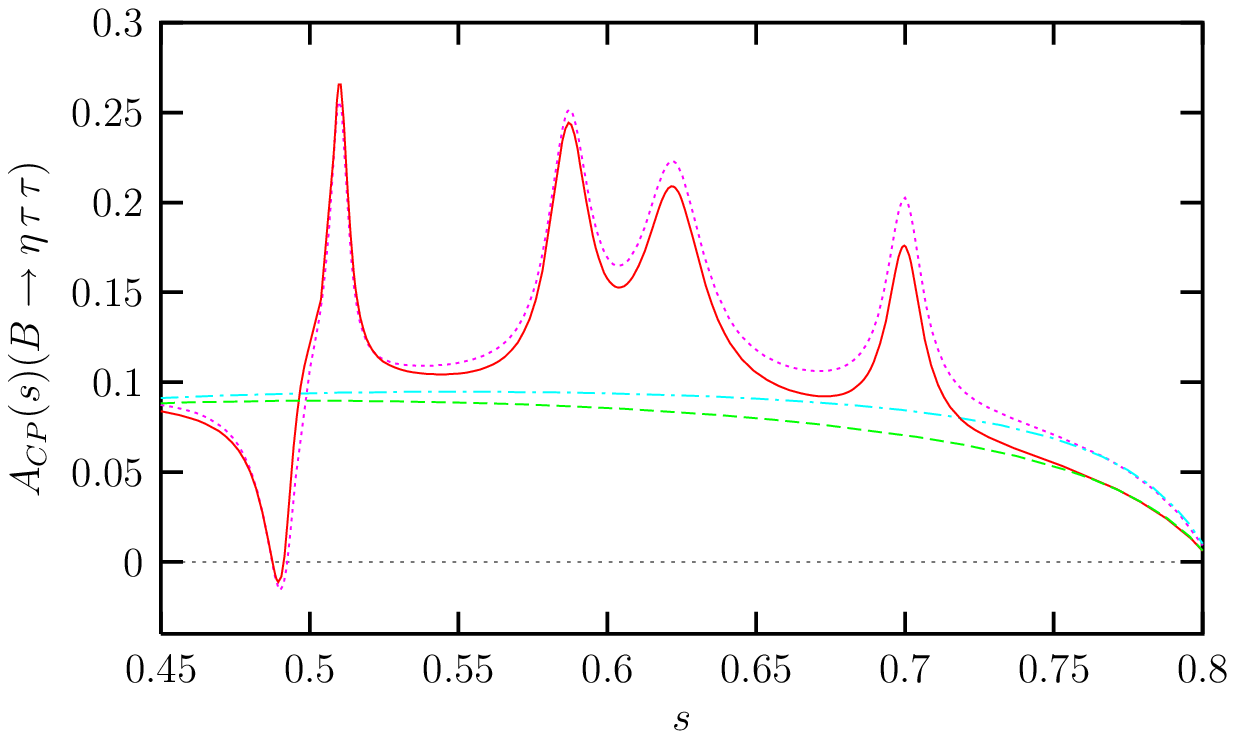} \vskip 0truein
\caption[]{$A_{CP}(s)$ for $B\rar \eta \, \tau^{+} \tau^{-}$ decay for  the parameter set-1 and set-3 
with (without) long distance contributions,  represented by the small dashed (dotted dashed) 
and the solid (dashed) curves, respectively.} \label{dACPetatau}
\end{figure}

\begin{figure}[htb]
\vskip 0truein \centering \epsfxsize=3.8in
\leavevmode\epsffile{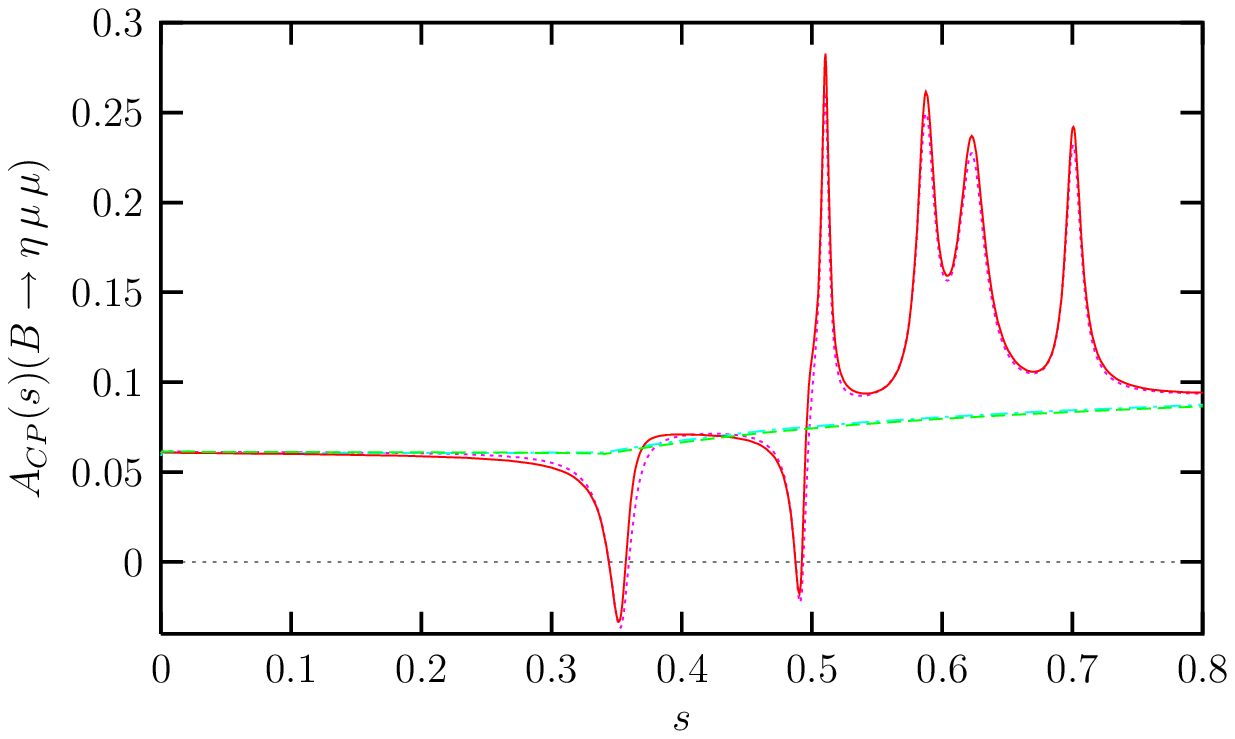} \vskip 0truein \caption[]{The same
as Fig.(\ref{dACPetatau}) but for the $B\rar \eta \, \mu^{+} \mu^{-}$ decay} \label{dACPetamu}
\end{figure}

\begin{figure}[htb]
\vskip 0truein \centering \epsfxsize=3.8in
\leavevmode\epsffile{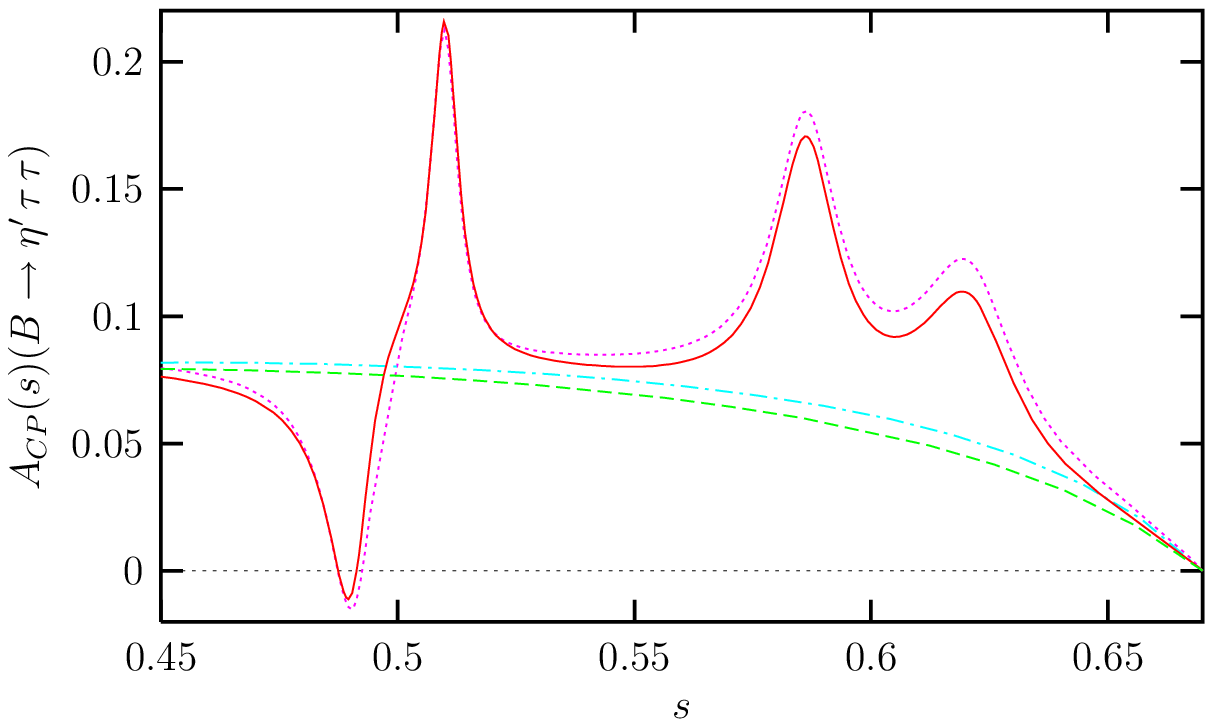} \vskip 0truein \caption[]{The same
as Fig.(\ref{dACPetatau}) but for the $B\rar \eta^{\prime} \, \tau^{+} \tau^{-}$ decay} \label{dACPetaptau}
\end{figure}

\begin{figure}[htb]
\vskip 0truein \centering \epsfxsize=3.8in
\leavevmode\epsffile{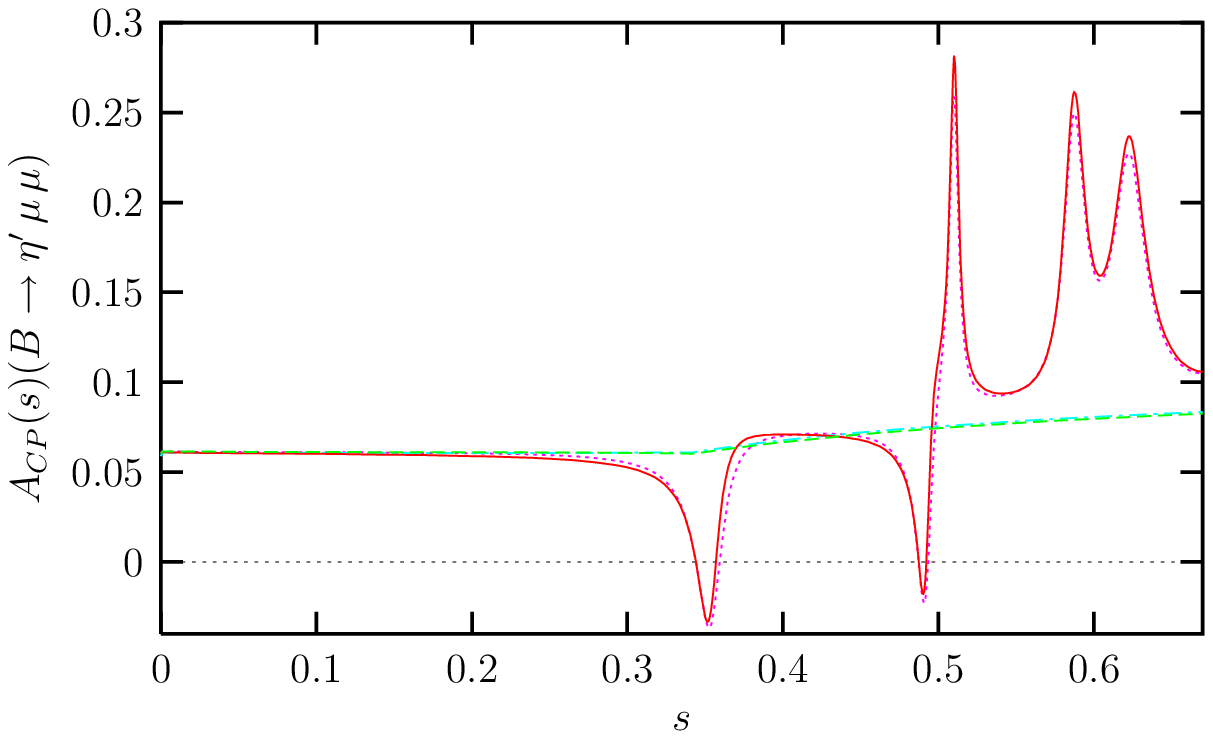} \vskip 0truein
\caption[]{The same
as Fig.(\ref{dACPetatau}) but for the $B\rar \eta^{\prime} \, \mu^{+} \mu^{-}$ decay} \label{dACPetapmu}
\end{figure}

Figs. (\ref{dACPetatau}) and  (\ref{dACPetamu}) are devoted to the $A_{CP}(s)$ as a function of $s$ for 
$B_{d}\rar \eta \, \tau^{+} \tau^{-}$ and $B_{d}\rar \eta \, \mu^{+} \mu^{-} $ decays, respectively.  In these figures, the small dashed (dotted dashed)  and the solid (dashed) curves represent the $A_{CP}(s)$ 
for  the parameter set-1 and set-3 with (without) long distance contributions.
The dependence of $A_{CP}$ on $s$  for the $\eta^{\prime}$ channel is plotted in Fig. (\ref{dACPetaptau}) and 
Fig. (\ref{dACPetapmu}), for $\ell=\tau$ and $\ell=\mu$, respectively.
We see from these figures that for $\ell=\mu$, $A_{CP}(s)$  is not very sensitive to the  choice of the 
parameters set-1 or set-3, reaching up to $28\%$  for the larger values of $s$ for 
both the $\eta$ and $\eta^{\prime}$ channels. However for $\ell=\tau$ case, $A_{CP}(s)$
gets slightly larger contribution from set-3 than set-1, but reaches at most $25\%$ in the 
small-$s$ region. We  note that $A_{CP}(s)$ is positive for all values of $s$, except in
some resonance regions. We also observe from table \ref{tab3} that including the long-distance
effects in calculating $<A_{CP}>$ changes the results only by $2-10\%$ for $\ell=\mu$ mode, but
for $\ell=\tau$, it becomes very sizable, $30-150\%$, depending on the sets of parameters used for
$(\rho ; \eta)$.

In conclusion, we have analyzed the \Bepell decays within the SM. We have found that,
 these decay modes have a significant $A_{CP}$, especially for $\ell=\tau$. Since  calculated $BR$s of 
 these decay modes are within the reach of forthcoming B-factories such as LHC-B, where approximately 
 $6\times 10^{11}$ $B_d$ mesons are expected to be produced per year, we may hope that it can be 
 measured in near future.  

\begin{table}
\begin{center}
\begin{tabular}{|c|c|c|c|c|c|}
 \hline
  $10 \cdot <A_{CP}>$  & {\scriptsize$\ell$} & set1          &  set2    & set3         &  set4 \\ \hline \hline
                       & $\tau$              & $1.291$       & $0.961$  & $2.271$      &  $0.840$    \\
		               &                     & $(0.899)$     & $(0.657)$& $(0.897)$    &  $(0.560)$ \\ \cline{2-6}
  $\eta$               & $\mu$               & $0.647$       & $0.496$  & $0.692$      &  $0.526$     \\
                       &                     & $(0.663)$     & $(0.484)$& $(0.671)$    &  $(0.490)$ \\ \cline{2-6}
                       & $e$                 & $0.647$       & $0.496$  & $0.693$      &  $0.526$    \\
					   &                     & $(0.663)$     & $(0.484)$& $(0.671)$    &  $(0.490)$  \\ \hline \hline
		               & $\tau$              & $0.926$       & $0.693$  & $0.886$      &  $0.656$    \\
		               &                     & $(0.699)$     & $(0.510)$& $(0.629)$    &  $(0.458)$ \\ \cline{2-6}
  $\eta'$              & $\mu$               & $0.578$       & $0.444$  & $0.593$      &  $0.452$     \\
                       &                     & $(0.637)$     & $(0.464)$& $(0.639)$    &  $(0.465)$ \\ \cline{2-6}
                       & $e$                 & $0.579$       & $0.444$  & $0.594$      &  $0.452$    \\
					   &                     & $(0.638)$     & $(0.464)$& $(0.640)$    &  $(0.465)$ \\ \hline
	\end{tabular}
  \end{center}
  \caption{The same as Table (\ref{tab2}), but for $<A_{CP}>$.}\label{tab3}
  \end{table}

\newpage
\end{document}